\documentclass[12pt]{article}

\usepackage[T1]{fontenc}
\usepackage[margin=1in]{geometry}
\usepackage{amsmath,amssymb,amsthm,mathtools,bm}
\usepackage{microtype,color}
\usepackage[hidelinks]{hyperref}

\newtheorem{theorem}{Theorem}[section]
\newtheorem{assumption}{Assumption}[section]
\newtheorem{proposition}[theorem]{Proposition}
\newtheorem{lemma}[theorem]{Lemma}

\newcommand{\E}{\mathbb{E}}
\newcommand{\Pp}{\mathbb{P}}
\newcommand{\Q}{\mathbb{Q}}
\newcommand{\F}{\mathcal{F}}

\newcommand{\sF}{\mathcal{F}}

\newcommand{\R}{\mathbb{R}}
\newcommand{\e}{\mathrm{e}}

\begin{document}

\begin{center}
{\Large Stock bubbles in limited participation models with power utilities}\footnote{ Kasper Larsen is corresponding author and has contact information: Email: \url{KL756@math.rutgers.edu} and mailing address: Department of Mathematics, Rutgers University, Hill Center 330 - Busch Campus, 110 Frelinghuysen Road, Piscataway, NJ 08854-8019, USA. This note is a response to a question asked during the author's presentation at the NUS QF Conference (2026). We thank Umut \c{C}etin, Paolo Guasoni, Nizar Touzi, and Marko Weber for raising and discussing this bubble problem. The author used ChatGPT to develop and check  mathematical arguments as well as to provide feedback on writing.}

\ \\

{\large \bf Kasper Larsen}\\
Rutgers University

\end{center}

\begin{center}
\today
\end{center}

\vspace{0.3cm}

\begin{abstract}
We prove that the equilibrium state-price deflated cum-dividend stock-price process in Guasoni, Larsen, and Leoni (2025) is a strict local martingale. Our result is based on proving that the unrestricted trader faces an endogenously complete model.
\end{abstract}

\section{Introduction and mathematical setting}

The limited participation model in  Basak and Cuoco (1998) gives equilibrium existence when both the unrestricted trader (can trade both the stock and the bank account) and the restricted trader (can only trade the bank account) have log utilities. Hugonnier  (2012) proves that the resulting equilibrium stock price is a bubble in the sense that the state-price-deflated cum-dividend stock price process is a strict local martingale. This notion of bubbles is standard, see, e.g.,  Loewenstein and Willard (2000), Cox and Hobson (2005), and Kardaras, Kreher, and Nikeghbali (2015). While the fundamental stock price is given as the state-price deflated future dividends, the equilibrium stock price is given by market clearing conditions, which give the equilibrium stock-price process as the aggregate wealth process.  Guasoni, Larsen, and Leoni (2025) prove equilibrium existence when both traders have the identical power utilities $\frac {x^{1-\gamma}}{1-\gamma}$ for $\gamma \in (0,1)$ and $x\ge0$. In this technical note, we prove that the equilibrium stock price in Guasoni, Larsen, and Leoni (2025) is also a bubble. As a by product,  we prove the resulting stock price volatility is non-singular, which renders the unrestricted trader's model endogenously complete.

We consider a one-point extension $\mathbb{R}_{\Delta}$ of $\R$, where $\Delta$ is a
cemetery state. We use the canonical path space $\Omega$ of paths that are c\`adl\`ag up to their lifetime and remain at
$\Delta$ thereafter. For $t\geq0$, let $B_t(\omega):=\omega(t)$ be the coordinate process, and let $\widetilde{\mathcal{F}}_t$ denote the
canonical filtration generated by $B$.  The argument of Lemma~2.6 in Kardaras, Kreher, and Nikeghbali
(2015) shows that $\widetilde{\sF}_t$ is a standard system. We let $\mathcal{F}_t$ denote the right-continuous extension of $\widetilde{\mathcal{F}}_t$.
We let $\mathbb{P}$ be the Wiener measure on $\bigvee_{t\ge0} \sF_t$, under which $B$ is a Brownian
motion. Proposition~2.8 and Theorem 2.12 in Kardaras, Kreher, and Nikeghbali (2015) allow us to use F\"ollmer measure extensions associated with strictly positive local martingales with respect to $\sF_t$.

\section{Main results}
As in Basak and Cuoco (1998), the exogenous dividend-rate process is a geometric Brownian motion given by
$$
dD_t = \mu_D D_t dt + \sigma_D D_t dB_t,\quad D_0>0,
$$
where $\mu_D\in\R$ and $\sigma_D,D_0>0$ are constants. We let $\gamma \in (0,1)$ be the common constant relative risk-aversion coefficient for both traders.  Throughout this paper, we use the equilibrium constructed in
Theorem~3.6 in Guasoni, Larsen, and Leoni (2025), which relies on the following parameter restrictions.

\begin{assumption}\label{ass_main} Let $D_0>0$, $\sigma_D>0$, $\gamma\in(0,1)$, $ \beta>0$,  and assume
\begin{equation}
    \varrho:=\beta-(1-\gamma)\mu_D
       +\frac12\gamma(1-\gamma)\sigma_D^2>0.
    \label{eq:standing-assumptions}
\end{equation}
\begin{flushright}
$\square$
\end{flushright}
\end{assumption}

\noindent Under this assumption, the existence proof and properties of the deterministic function $h \in \mathcal{C}^1([0,1])$ can be found in Guasoni, Larsen, and Leoni (2025). The specific ODE for $h$ is not used here but $h$ is valued in $[\gamma,1]$ with $h(0)=\gamma$ and $ h(1)=1$. The following analysis uses the state-process $Y$ valued in $(0,1)$ with the autonomous dynamics\footnote{ The SDE for $Y$ has a global pathwise unique strong solution. In equilibrium, the initial condition $Y_0$  is determined by the traders' endowed holdings of the bank account.}
\begin{align} \label{eq:muY_sigmaY}
\begin{split}
dY_t &= \mu_Y(Y_t) dt + \sigma_Y(Y_t) dB_t,\quad Y_0\in(0,1),\\
    \mu_Y(y):       &=\frac{\sigma_D^2(1-y)
        \bigl(1+\gamma+2\gamma yh(y)-2y(1+\gamma)\bigr)}
        {2y h(y)^2},   \quad \sigma_Y(y):= \sigma_D\frac{1-y}{h(y)}, \quad y\in (0,1),
\end{split}
\end{align}
where the drift and volatility functions are taken from Eqs. (3.1)--(3.2) in Guasoni, Larsen, and Leoni (2025).

The equilibrium bank account and stock price processes have dynamics 
\begin{align}
    \frac{d S^{(0)}_t}{S^{(0)}_t}
        &= r(Y_t)dt,
        \qquad S^{(0)}_0=1,                                      \label{eq:S0}\\
    d S_t
        &= r(Y_t)S_tdt
        +S_t\sigma_{S,t}\bigl(\kappa(Y_t)dt+dB_t\bigr)
        -D_tdt,                                               \label{eq:S}
\end{align}
where the market price of risk $\kappa$ and interest rate $r$ are from Eq. (3.26) in Guasoni, Larsen, and Leoni (2025) and read
\begin{equation}\label{kappa_r}
    \kappa(y)
      :=\gamma\sigma_D
        \left(\frac{1-y}{y h(y)}+1\right),\quad  r(y):=
\varrho+\mu_D
-\gamma\sigma_D^2
-\frac{\gamma(1+\gamma)\sigma_D^2(1-y)}
{2y h(y)^2},
\quad y\in(0,1).
\end{equation}

Our first result is:
\begin{proposition}\label{prop:sigma-positive} Under Assumption \ref{ass_main},  the volatility coefficient in \eqref{eq:S} has the form $\sigma_{S,t}=\sigma_S(Y_t)$ with $ \sigma_S:(0,1) \to (0,\infty)$ continuous.

\end{proposition}

Guasoni, Larsen, and Leoni (2025) show in their Eq. (3.27) that the unrestricted trader 1 and the restricted trader 2 have equilibrium consumption rates given by
\begin{equation}
    \widehat c_{1,t}=D_tY_t,
    \qquad
    \widehat c_{2,t}=D_t(1-Y_t).
    \label{eq:consumption}
\end{equation}
Based on first-order conditions, the two optimal state-price deflators (dual optimizers) are
\begin{equation}
    \widehat Z_{1,t}
    =\e^{-\beta t}(D_tY_t)^{-\gamma},
    \qquad
    \widehat Z_{2,t}
    =\e^{-\beta t}\bigl(D_t(1-Y_t)\bigr)^{-\gamma}.
    \label{eq:SPDs}
\end{equation}
It\^o's lemma produces the dynamics 
\begin{equation}
    \frac{d \widehat Z_{1,t}}{\widehat Z_{1,t}}
      =-r(Y_t)dt-\kappa(Y_t)dB_t,
    \qquad
    \frac{d \widehat Z_{2,t}}{\widehat Z_{2,t}}
      =-r(Y_t)dt-\nu(Y_t)dB_t,
    \label{eq:Z-dynamics}
\end{equation}
where $\kappa$ and $r$ are as in \eqref{kappa_r} and 
\begin{align}        \label{eq:nu}
   \nu(y)
      :=-\gamma\sigma_D\frac{1-h(y)}{h(y)},\quad y\in[0,1].
\end{align}

The next result shows that no matter what state-price deflator (dual element) is used for deflating future dividends, the cum dividend stock-price process will never be a martingale. In other words, similar to Hugonnier  (2012) proving the log-log model in Basak and Cuoco (1998) has a bubble, the equilibrium stock-price process \eqref{eq:S} for the identical power-power model also has a bubble.

\begin{theorem}\label{thm:no-nu} Under Assumption \ref{ass_main}, there is no  \(\eta\in \mathcal L^2_{\mathrm{loc}}\) such that the process
\begin{equation}
  \frac{d\widehat Z_t^\eta}{\widehat Z_t^\eta}
  =-r(Y_t)dt-\eta_tdB_t,
  \qquad \widehat Z_0^\eta>0,
  \label{eq:spd}
\end{equation}
produces a martingale of 
\begin{equation}
  \widehat Z_t^\eta S_t
  +\int_0^t\widehat Z_s^\eta D_sd s.
  \label{eq:main-conclusion-spd}
\end{equation}

\end{theorem}

As a by product, the proof shows that only $\eta_t = \kappa(Y_t)$ with $\kappa$ as in \eqref{kappa_r}, produces the local martingale property of \eqref{eq:main-conclusion-spd}. Even though  $\eta_t = \kappa(Y_t)$ does not produce the martingale property \eqref{eq:main-conclusion-spd}, this property is sufficient to prove that the unrestricted trader 1 faces an endogenously complete market.

\section{Proofs}

\subsection{Proof of Proposition \ref{prop:sigma-positive}}

\noindent \textbf{Step 1/3:} The martingale property of $\mathcal E \big((1-\gamma)\sigma_D\cdot B\big)$ allows us to use 
 Theorem~2.12 in Kardaras, Kreher, and Nikeghbali (2015) to construct a probability measure $\Q$ on
$\bigvee_{t\ge0}\mathcal F_t$ such that  on every finite horizon \(T \in (0,\infty)\), we have
\[
  \left.\frac{d\Q}{d\Pp}\right|_{\F_T}
  =\exp\!\left(
      (1-\gamma)\sigma_DB_T
      -\frac12(1-\gamma)^2\sigma_D^2T
    \right).
\]
Under \(\Q\), Girsanov's theorem gives the dynamics of \eqref{eq:muY_sigmaY} to be
\begin{align}\label{dY_underQ}
  d Y_t
  = \mu^\Q_Y(Y_t)dt
   +\sigma_Y(Y_t)dB_t^{\Q},\quad  \mu^\Q_Y(y)
  :=\mu_Y(y)+(1-\gamma)\sigma_D\sigma_Y(y),\quad y \in (0,1).
\end{align}
For $T>0$, the equivalence of $\Pp$ and $\Q$ on 
$\sF_T$ gives $\Q(Y_s\in (0,1)\forall s\in [0,T])=1$. We define
\begin{align}\label{kappa}
  k(y):=
  \E^{\Q}\!\left[
    \int_0^\infty e^{-\varrho s}Y_s^{1-\gamma}d s
  \right],
  \qquad
  v(y):=y^\gamma k(y),\quad Y_0=y\in(0,1).
\end{align}
Because  $Y_s\in (0,1)$ for $ s>0$ and \(\varrho>0\), we have
\[
  0<k(y)<\frac1{\varrho},
  \qquad
  v(y)>0,\quad y \in (0,1).
\]
Because the drift and volatility coefficients of \(Y\) in  \eqref{dY_underQ} are locally Lipschitz on
\((0,1)\), pathwise uniqueness holds for \eqref{dY_underQ}. Continuity and pathwise uniqueness imply the one-dimensional
comparison property in that
\[
y_1\le y_2 \Rightarrow \Q(  Y_s^{y_1}\leq Y_s^{y_2} \;\forall s\geq0)=1.
\]
Therefore, $k$ in \eqref{kappa} is non-decreasing and the two limits $k(0+):= \lim_{y\downarrow0}k(y)$ and  $k(1-):= \lim_{y\uparrow1}k(y)$ exist. 
Because \(k>0\) in \eqref{kappa} and \(y\mapsto y^\gamma\) is strictly increasing, \(v(y)\) is strictly increasing:
\[
y_1< y_2 \Rightarrow \  v(y_1)=y_1^\gamma k(y_1)
  \leq y_1^\gamma k(y_2)
  <y_2^\gamma k(y_2)=v(y_2).
\]

\noindent \textbf{Step 2/3:} Fix $0<a<b<1$, and let
\[
  \tau_{a,b}:=\inf\{t\geq0:Y_t\notin(a,b)\}.
\]
The functions $\mu_Y^\Q$ and $\sigma_Y$ are continuously differentiable
on $(0,1)$, and $\sigma_Y$ is bounded away from zero on every compact
subinterval of $(0,1)$, hence, $Y$ is a regular diffusion on $(0,1)$. Lemma~33.8 in Kallenberg
(2021) gives
\[
  \sup_{y\in[a,b]}\E^\Q[\tau_{a,b}]<\infty.
\]
Therefore, we have
\[
  \Q(\tau_{a,b}<\infty)=1,
  \qquad y\in(a,b).
\]

The Dynkin operator associated with \eqref{dY_underQ} is
\[
  \mathcal L^{\Q}f(y)
  :=\frac12\sigma_Y(y)^2f''(y)+\mu_Y^\Q(y)f'(y),\quad y\in (0,1),\quad f \in C^2((0,1)).
\]
We consider the Dirichlet problem
\begin{equation}
  \begin{cases}
    \mathcal L^{\Q}u(y)-\varrho u(y)=-y^{1-\gamma},
      &y\in(a,b),\\
    u(a)=k(a),\qquad u(b)=k(b).
  \end{cases}
  \label{eq:k-local-Dirichlet}
\end{equation}
Theorem~6.14 in Gilbarg and Trudinger (2001) gives a unique solution
\(u_{a,b}\in C^2((a,b))\cap C([a,b])\). Applying It\^o's formula to
$e^{-\varrho(t\wedge\tau_{a,b})}
u_{a,b}(Y_{t\wedge\tau_{a,b}})$ and using
$\Q(\tau_{a,b}<\infty)=1$ together with Lebesgue's convergence theorems gives (this is  Exercise~9.12 in \O ksendal, 2003) 
\begin{equation}
  u_{a,b}(y)
  =\E^{\Q}\!\left[
      \int_0^{\tau_{a,b}}e^{-\varrho t}Y_t^{1-\gamma}dt
      +e^{-\varrho\tau_{a,b}}k(Y_{\tau_{a,b}})
    \right].
  \label{eq:k-local-FK}
\end{equation}
To see that the strong Markov property
applied to the defining expectation \eqref{kappa} for \(k\) gives the right-hand
side of \eqref{eq:k-local-FK}, we define the integrand in \eqref{kappa} as
\[
  \Psi
  :=
  \int_0^\infty e^{-\varrho s}Y_s^{1-\gamma}d s.
\]
For $t>0$, let $\theta_t$ denote the shift operator for the Markov
process $Y$. Then, Eq. (9.3.7) in \O ksendal (2003) gives
\begin{align}\label{shift_operator1}
 \theta_t \Psi
   =
  \int_0^\infty e^{-\varrho u}Y_{u+t}^{1-\gamma}d u
  =
  \int_t^\infty
    e^{-\varrho(s-t)}Y_s^{1-\gamma}d s,
\end{align}
where the last equality follows from the change of variables $s=u+t$.  On the set $\{\tau_{a,b}<\infty\}$, this produces
\begin{align}\label{Psi_new}
  \Psi
  =
  \int_0^{\tau_{a,b}}
    e^{-\varrho s}Y_s^{1-\gamma}d s
  +
  e^{-\varrho\tau_{a,b}}
  \theta_{\tau_{a,b}}\Psi.
\end{align}
The strong Markov property gives
\[
  \E^\Q\!\left[
   \theta_{\tau_{a,b}} \Psi
    \,\middle|\,\mathcal F_{\tau_{a,b}}
  \right]
  =  \E^\Q\!\left[
    \Psi
  \right]\Big|_{y=Y_{\tau_{a,b}}}
=
  k(Y_{\tau_{a,b}}).
\]
Using this when taking expectations in \eqref{Psi_new} gives
\[
  k(y)
  =
  \E^\Q\!\left[
      \int_0^{\tau_{a,b}}e^{-\varrho s}Y_s^{1-\gamma}d s
      +e^{-\varrho\tau_{a,b}}k(Y_{\tau_{a,b}})
  \right]
  =
  u_{a,b}(y).
\]
This $C^2$ property together with existence of $k(0+)$ and  $k(1-)$ from the previous step gives  $k\in C([0,1])$. All in all, $ k\in C^2((0,1))\cap C([0,1])$,  $k(y) \in (0, \frac1{\varrho})$ for all $y \in (0,1)$, $k$ is non-decreasing, and $k$ satisfies the ODE
\begin{equation}
  \frac12\sigma_Y(y)^2k''(y)
  +\mu_Y^\Q(y)k'(y)
  -\varrho k(y)=-y^{1-\gamma},
  \qquad y\in(0,1).
  \label{eq:k-ODE}
\end{equation}
Consequently, $v$ from \eqref{kappa} satisfies \(v\in C^2((0,1))\) and
\[
  v'(y)
  =\gamma y^{\gamma-1}k(y)+y^\gamma k'(y)
  \geq \gamma y^{\gamma-1}k(y)>0,
  \qquad y\in(0,1).
\]

\noindent\textbf{Step 3/3:} Let $Y_0 \in (0,1)$ be the equilibrium initial state-process value 
 determined by the traders' initial bank endowments.  Eq. (4.18) of Guasoni, Larsen, and Leoni (2025) gives
\begin{align}\label{wealth1}
\begin{split}
  \widehat X_{1,t}
  &=\frac{1}{\widehat Z_{1,t}}
    \E_t\!\left[
      \int_t^\infty \widehat c_{1,u}\widehat Z_{1,u}d u
    \right] \\
  &=e^{\beta t}(D_tY_t)^\gamma
    \E_t\!\left[
      \int_t^\infty e^{-\beta u}(D_uY_u)^{1-\gamma}d u
    \right] \\
  &=D_tY_t^\gamma
    \E_t\!\left[
      \int_0^\infty
      e^{-\beta s}
      \left(\frac{D_{t+s}}{D_t}\right)^{1-\gamma}
      Y_{t+s}^{1-\gamma}d s
    \right].
    \end{split}
\end{align}
Moreover,
\[
  e^{-\beta s}
  \left(\frac{D_{t+s}}{D_t}\right)^{1-\gamma}
  =e^{-\varrho s}
   \frac{L_{t+s}}{L_t},
\quad   L_t
  :=\exp\!\left(
      (1-\gamma)\sigma_DB_t
      -\frac12(1-\gamma)^2\sigma_D^2t
    \right).
\]
Because the integrand is positive, the conditional Tonelli's theorem gives
\begin{align*}
  D_tY_t^\gamma
  \E_t\!\left[
    \int_0^\infty
      e^{-\varrho s}
      \frac{L_{t+s}}{L_t}
      Y_{t+s}^{1-\gamma}d s
  \right] 
  &=
  D_tY_t^\gamma
  \int_0^\infty e^{-\varrho s}
  \E_t\!\left[
      \frac{L_{t+s}}{L_t}
      Y_{t+s}^{1-\gamma}
  \right]d s.
\end{align*}
For every $s\geq0$, Bayes' formula and the  Markov property under $\Q$ give
\[
  \E_t\!\left[
      \frac{L_{t+s}}{L_t}
      Y_{t+s}^{1-\gamma}
  \right]
  =
  \E_t^\Q\!\left[
      Y_{t+s}^{1-\gamma}
      \right]=
  \left.
  \E^\Q\!\left[Y_s^{1-\gamma}\right]
  \right|_{Y_0=Y_t}.
\]
Consequently, the representation in \eqref{wealth1} and  the definition
of $k$ in \eqref{kappa} give 
\begin{align*}
  \widehat X_{1,t}
  &=
  D_tY_t^\gamma
  \left.
  \E^\Q\!\left[
    \int_0^\infty
      e^{-\varrho s}Y_s^{1-\gamma}d s
  \right]
  \right|_{Y_0=Y_t} =
  D_tY_t^\gamma k(Y_t)
  =
  D_tv(Y_t).
\end{align*}
Combining this with the first identity in (4.18) of Guasoni, Larsen, and Leoni (2025)  gives 
\begin{align}\label{S_equilibrium}
  S_t=\widehat X_{1,t}+\widehat X_{2,t} = D_tv(Y_t) + D_t g(Y_t)(1-Y_t)^\gamma,
\end{align}
where the second equality uses (4.15) of Guasoni, Larsen, and Leoni (2025) and $g$ solves the ODE in Theorem 3.6(1) in  Guasoni, Larsen, and Leoni (2025).   Because trader 2 cannot hold the stock, the $d B$ coefficient in $d S$ equals the $d B$ coefficient in $d D_tv(Y_t)$.
 It\^o's lemma produces the coefficient in \eqref{eq:S} to be
\begin{align*}
  S_t\sigma_{S,t}
=D_t\Bigl(
      \sigma_Dv(Y_t)+\sigma_Y(Y_t)v'(Y_t)
    \Bigr).
\end{align*}
By dividing by $S$ and using \eqref{S_equilibrium}, we see
\begin{align}\label{sigma_S}
  \sigma_S(y)=
  \frac{
    \sigma_Dv(y)+\sigma_Y(y)v'(y)
  }{
    v(y)+g(y)(1-y)^\gamma
  },
  \qquad y\in(0,1).
\end{align}
Now \(\sigma_D>0\), \(v(y)>0\), \(\sigma_Y(y)>0\), and \(v'(y)>0\).  Therefore,
\[
  \sigma_Dv(y)+\sigma_Y(y)v'(y)
  >\sigma_Dv(y)>0.
\]
Because $v(y)>0$, $g(y)>0$, and $(1-y)^\gamma>0$, also the denominator in \eqref{sigma_S} is strictly positive.

Since $v\in C^2((0,1))$, $\sigma_Y$ and $g$ are continuous on
$(0,1)$, and the denominator in \eqref{sigma_S} is strictly positive,
it follows that $\sigma_S$ is continuous.

\begin{flushright}
$\square$
\end{flushright}

\subsection{Density ratio and measure changes}

Unlike in Basak and Cuoco (1998), the ratio of the processes in \eqref{eq:SPDs} given by $\Lambda_t    :=\frac{\widehat Z_{1,t}}{\widehat Z_{2,t}}
    =\left(\frac{1-Y_t}{Y_t}\right)^{\gamma}$ is not
a local martingale under $\Pp$.\footnote{While we do not need it, the ratio is a nonnegative local submartingale under $\Pp$.} Because $h(y) \in [\gamma,1]$, we get the constant upper bound for \eqref{eq:nu} given by
\begin{equation}
    |\nu(y)|\leq \sigma_D(1-\gamma),
    \qquad y\in[0,1].
    \label{eq:nu-bounded}
\end{equation}
This bound ensures the stochastic exponential $L_{2,t} := \mathcal{E} (-\nu \cdot B)_t$ is a martingale. Theorem~2.12 in Kardaras, Kreher, and Nikeghbali (2015) gives a locally equivalent probability
measure $\Q_2$ with Radon-Nikodym derivative $\frac{d\Q_2}{d\Pp}=L_{2,t}$ on $\sF_t$ for all $t\ge0$.  Girsanov's theorem ensures 
\begin{equation}
    W_{2,t}:=B_t+\int_0^t\nu(Y_s)d s
    \label{eq:W2}
\end{equation}
 is a $\Q_2$-Brownian motion.  Moreover, It\^o's lemma and  \eqref{eq:W2} give
\begin{align}    \label{eq:Lambda-P}
    \frac{d\Lambda_t}{\Lambda_t}
      &=\frac{\gamma^2\sigma_D^2(1-h(Y_t))}
              {Y_t h(Y_t)^2}dt
        -\frac{\gamma\sigma_D}{Y_t h(Y_t)}dB_t
      =-\frac{\gamma\sigma_D}{Y_t h(Y_t)}d W_{2,t}.
\end{align}
This shows that $\Lambda$ is a positive local martingale under $\Q_2$ and, in particular, a supermartingale.
\begin{lemma}Under Assumption \ref{ass_main}, the ratio $\Lambda$ is a strict local martingale under $\Q_2$ and 
trader 1's normalized discounted state-price density 
\begin{equation}
    L_{1,t}
      :=\frac{S^{(0)}_t\widehat Z_{1,t}}{\widehat Z_{1,0}},
    \label{eq:L-def}
\end{equation}
is a strict local martingale under $\Pp$.
\end{lemma}
\begin{proof} We first consider  $\Lambda_t        =\left(\frac{1-Y_t}{Y_t}\right)^{\gamma}$. Its inverse is given by
\begin{equation}
    Y_t=y(\Lambda_t),
    \qquad
    y(\lambda):=\frac{1}{1+\lambda^{1/\gamma}},\quad \lambda \in (0,\infty).
    \label{eq:y-lambda}
\end{equation}
This allows us to rewrite the dynamics \eqref{eq:Lambda-P} as
 an autonomous diffusion with dynamics
\begin{equation}
    d\Lambda_t=-\Lambda_t\frac{\gamma\sigma_D}{Y_t h(Y_t)}d W_{2,t}= -a(\Lambda_t)d W_{2,t},
    \qquad
    a(\lambda)
      :=\lambda\frac{\gamma\sigma_D}
              {y(\lambda)h(y(\lambda))},\quad \lambda \in (0,\infty).
    \label{eq:a-lambda}
\end{equation}
To use Delbaen and Shirakawa (2002), we normalize and set $N_t:=\Lambda_t/\Lambda_0$ and $\widetilde W:=-W_2$ to get
\[
  d N_t=\widetilde a(N_t)d\widetilde W_t,
  \qquad N_0=1,\qquad   \widetilde a(x):=\frac{a(\Lambda_0x)}{\Lambda_0}, \qquad x>0.
\]
Theorem~1.6 in Delbaen and Shirakawa (2002) shows that $N$ is a
martingale if and only if
\begin{equation}
  \int^\infty\frac{x}{\widetilde a(x)^2}d x=  \int^\infty\frac{\lambda}{a(\lambda)^2}d\lambda=\infty,
    \label{eq:DS-test}
\end{equation}
where we used the change of variables $\lambda:=\Lambda_0x$. To see that the second integral in \eqref{eq:DS-test} is finite, we use \eqref{eq:y-lambda} to see
$$
  \lim_{\lambda\uparrow \infty}  y(\lambda)\lambda^{1/\gamma}
  =
  \lim_{\lambda\uparrow \infty} \frac{\lambda^{1/\gamma}}{1+\lambda^{1/\gamma}}
=
  \lim_{\lambda\uparrow \infty} \frac{1}{1+\lambda^{-1/\gamma}}=1.
$$
Because $\lim_{\lambda\uparrow \infty} h\big(y(\lambda)\big)=h(0)=\gamma$, \eqref{eq:a-lambda} gives
\[
  \lim_{\lambda\uparrow \infty} \frac{a(\lambda)}{\lambda^{1+1/\gamma}}
=
  \lim_{\lambda\uparrow \infty} \frac{\gamma\sigma_D}
{\lambda^{1/\gamma}y(\lambda)h(y(\lambda))}=\sigma_D.
\]
Therefore, up to an irrelevant constant, the integrand in \eqref{eq:DS-test} behaves asymptotically as 
$\lambda^{-1-2/\gamma}$ and so \eqref{eq:DS-test} fails.

Second, we consider $L_{1}$ defined in \eqref{eq:L-def}. We have
\[
    \E^\Pp[L_{1,t}]
    =
    \E^\Pp\left[
        L_{2,t}\frac{\Lambda_t}{\Lambda_0}
    \right]
    =
    \E^{\Q_2}\left[
        \frac{\Lambda_t}{\Lambda_0}
    \right].
\]
Since $\Lambda$ is a positive strict local martingale under $\Q_2$, the
right-hand side is strictly smaller than one for some $t>0$.
Because $L_1$ is a positive local martingale under $\Pp$, it follows
that $L_1$ is a strict local martingale under $\Pp$.

\end{proof}

\subsection{Proof of Theorem \ref{thm:no-nu}}

We argue by contradiction and suppose the
process in \eqref{eq:main-conclusion-spd} is a martingale for some $\eta\in\mathcal L^2_{\mathrm{loc}}$. It\^o's
lemma, \eqref{eq:S}, and \eqref{eq:spd} give
\begin{align*}
  d\left(
    \widehat Z_t^\eta S_t
    +\int_0^t\widehat Z_s^\eta D_sd s
  \right)
  &=
  \widehat Z_t^\eta S_t\sigma_S(Y_t)
  \bigl(\kappa(Y_t)-\eta_t\bigr)dt \\
  &\quad+
  \widehat Z_t^\eta S_t
  \bigl(\sigma_S(Y_t)-\eta_t\bigr)dB_t .
\end{align*}
By being a local martingale, the finite-variation part must
vanish. Since $\widehat Z^\eta>0$, $S>0$, and
Proposition~\ref{prop:sigma-positive} gives $\sigma_S(Y_t)>0$, we get
\[
  \eta_t=\kappa(Y_t),
  \qquad
  dt\otimes d \Pp\text{-a.e.}
\]
Therefore, $\widehat Z^\eta$ is a positive constant multiple of
$\widehat Z_1$. So it suffices to show that
\[
  M_t
  :=
  \widehat Z_{1,t}S_t
  +\int_0^t\widehat Z_{1,s}D_sd s
\]
cannot be a martingale.

Set
\[
  \overline S_t:=\frac{S_t}{S_t^{(0)}},
  \qquad
  \overline X_{2,t}:=
  \frac{\widehat X_{2,t}}{S_t^{(0)}}.
\]
Because trader 2 can trade only the bank account,
\[
  d\overline X_{2,t}
  =
  -\frac{\widehat c_{2,t}}{S_t^{(0)}}dt
  =
  -\delta_2(Y_t)\overline X_{2,t}dt,
\]
where, using the representation of $g$ in Theorem 3.6(1) in Guasoni, Larsen, and Leoni
(2025),
\[
\delta_2(y)
:=\frac{(1-y)^{1-\gamma}}{g(y)}
=\frac{1}{g(0)}
\exp\left(\int_0^y\frac{h(q)-1}{1-q}d q\right)
\leq\frac{1}{g(0)}.
\]
Consequently, for every $T>0$,
\[
  \overline X_{2,t}
=\overline X_{2,0}
\exp\left(-\int_0^t\delta_2(Y_s)d s\right)
\geq \overline X_{2,0}e^{-T/g(0)}
=:c_T>0,\qquad 0\leq t\leq T.
\]
Since $S=\widehat X_1+\widehat X_2$, we have
\[
  \overline S_t\geq\overline X_{2,t}\geq c_T,
  \qquad 0\leq t\leq T.
\]

Using \eqref{eq:L-def},
\[
  \frac{M_t}{\widehat Z_{1,0}}
  =
  L_{1,t}\overline S_t
  +\int_0^t
    L_{1,s}\frac{D_s}{S_s^{(0)}}d s
  \geq
  c_TL_{1,t},
  \qquad 0\leq t\leq T.
\]
If $M$ were a martingale, then, on every finite interval $[0,T]$,
$M/\widehat Z_{1,0}$ would be a nonnegative martingale and hence of
class $D$. Therefore, for every stopping time $\tau\leq T$,
\[
  0\leq L_{1,\tau}
  \leq
  \frac{1}{c_T\widehat Z_{1,0}}M_\tau.
\]
It follows that $L_1$ is of class $D$ on $[0,T]$. Since $L_1$ is a
local martingale, $L_1$ is  therefore a martingale on $[0,T]$.
As $T>0$ is arbitrary, $L_1$ would is a  martingale under $\Pp$, but that 
contradicts Lemma~3.1.
\begin{flushright}
$\square$
\end{flushright}

\end{document}